\input{aipcheck}

\documentclass[
    ,final            
  ]
  {aipproc}

\layoutstyle{8x11single}

\begin{document}

\title{Improved Collective Thomson Scattering
 measurements of fast ions at ASDEX Upgrade}


\author{J. Rasmussen}{
  address={Association Euratom-DTU, Technical University of Denmark, Department of Physics,
DTU Ris{\o} Campus, DK-4000 Roskilde, Denmark}
}

\author{S. K. Nielsen}{
  address={Association Euratom-DTU, Technical University of Denmark, Department of Physics,
DTU Ris{\o} Campus, DK-4000 Roskilde, Denmark}
}

\author{M. Stejner}{
  address={Association Euratom-DTU, Technical University of Denmark, Department of Physics,
DTU Ris{\o} Campus, DK-4000 Roskilde, Denmark}
}

\author{M. Salewski}{
  address={Association Euratom-DTU, Technical University of Denmark, Department of Physics,
DTU Ris{\o} Campus, DK-4000 Roskilde, Denmark}
}

\author{A. S. Jacobsen}{
  address={Association Euratom-DTU, Technical University of Denmark, Department of Physics,
DTU Ris{\o} Campus, DK-4000 Roskilde, Denmark}
}

\author{S.~B.~Korsholm}{
  address={Association Euratom-DTU, Technical University of Denmark, Department of Physics,
DTU Ris{\o} Campus, DK-4000 Roskilde, Denmark}
}

\author{F. Leipold}{
  address={Association Euratom-DTU, Technical University of Denmark, Department of Physics,
DTU Ris{\o} Campus, DK-4000 Roskilde, Denmark}
}

\author{F. Meo}{
  address={Association Euratom-DTU, Technical University of Denmark, Department of Physics,
DTU Ris{\o} Campus, DK-4000 Roskilde, Denmark}
}

\author{P. K. Michelsen}{
  address={Association Euratom-DTU, Technical University of Denmark, Department of Physics,
DTU Ris{\o} Campus, DK-4000 Roskilde, Denmark}
}

\author{D. Moseev}{
  address={Association Euratom-FOM Institute DIFFER, 3430 BE Nieuwegein, The Netherlands}
}

\author{M.~Schubert}{
  address={Max-Planck-Institut f\"ur Plasmaphysik, EURATOM-Association, Boltzmannstr.~2, 85748 Garching, Germany}
}

\author{J. Stober}{
  address={Max-Planck-Institut f\"ur Plasmaphysik, EURATOM-Association, Boltzmannstr.~2, 85748 Garching, Germany}
}

\author{G. Tardini}{
  address={Max-Planck-Institut f\"ur Plasmaphysik, EURATOM-Association, Boltzmannstr.~2, 85748 Garching, Germany}
}

\author{D. Wagner}{
  address={Max-Planck-Institut f\"ur Plasmaphysik, EURATOM-Association, Boltzmannstr.~2, 85748 Garching, Germany}
}

\author{the ASDEX Upgrade Team}{
  address={Max-Planck-Institut f\"ur Plasmaphysik, EURATOM-Association, Boltzmannstr.~2, 85748 Garching, Germany}
}

\begin{abstract}
Understanding the behaviour of the confined fast ions is important in both current and future fusion experiments. These ions play a key role in heating the plasma  and will be crucial for achieving conditions for burning plasma in next-step fusion devices. Microwave-based Collective Thomson Scattering (CTS) is well suited for reactor conditions and offers such an opportunity by providing measurements of the confined fast-ion distribution function resolved in space, time and 1D velocity space. 
We currently operate a CTS system at ASDEX Upgrade using a gyrotron which generates probing radiation at 105 GHz.  
A new setup using two independent receiver systems has enabled improved subtraction of the background signal,
and hence the first accurate characterization of fast-ion properties.
Here we review this new dual-receiver CTS setup
and present results on fast-ion measurements based on the improved background characterization.
These results have been obtained both with and without NBI heating, and with the
measurement volume located close to the centre of the plasma. 
The measurements agree  quantitatively with predictions of numerical simulations. Hence, CTS studies of
fast-ion dynamics at ASDEX Upgrade are now feasible.
The new background subtraction technique could be important for 
the design of 
CTS systems in other fusion experiments.
\end{abstract}

\maketitle

\section{Introduction: The new dual-receiver setup at ASDEX Upgrade}

Collective Thomson Scattering (CTS) is based on the scattering of electromagnetic waves
off microscopic fluctuations in the plasma. In mm-wave CTS, an incident probe
beam with wave vector ${\bf k}^i$ scatters off ion-driven collective fluctuations
in the electron distribution
 with wave vector ${\bf k}^\delta$. A receiver antenna accepts a narrow beam of the scattered 
radiation for which ${\bf k}^s = {\bf k}^i + {\bf k}^\delta$ 
(Fig.~\ref{fig,geom}). For ions moving with velocity ${\bf v_i}$, 
the scattering signal will be frequency-shifted by
$\omega^\delta \approx {\bf v_i} \cdot {\bf k}^\delta$ and thus contains
information about the ion velocity distribution projected along ${\bf
  k}^\delta$. CTS can resolve this scattering signal in both
time and space. From this, the distribution function of
thermal (bulk) and non-thermal (fast) ions can be inferred, along with
the temperature, density and rotation velocity of bulk ions \citep{kors06}.

\begin{figure}
\includegraphics[height=5.4cm]{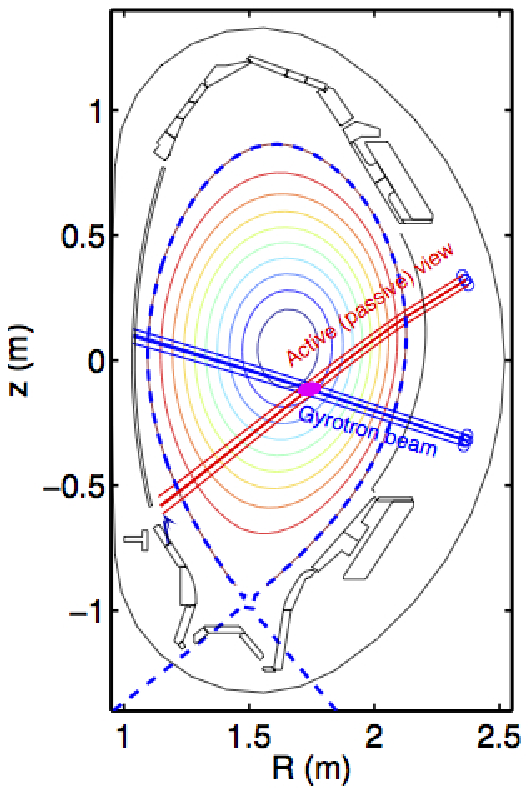}
\includegraphics[height=5.4cm]{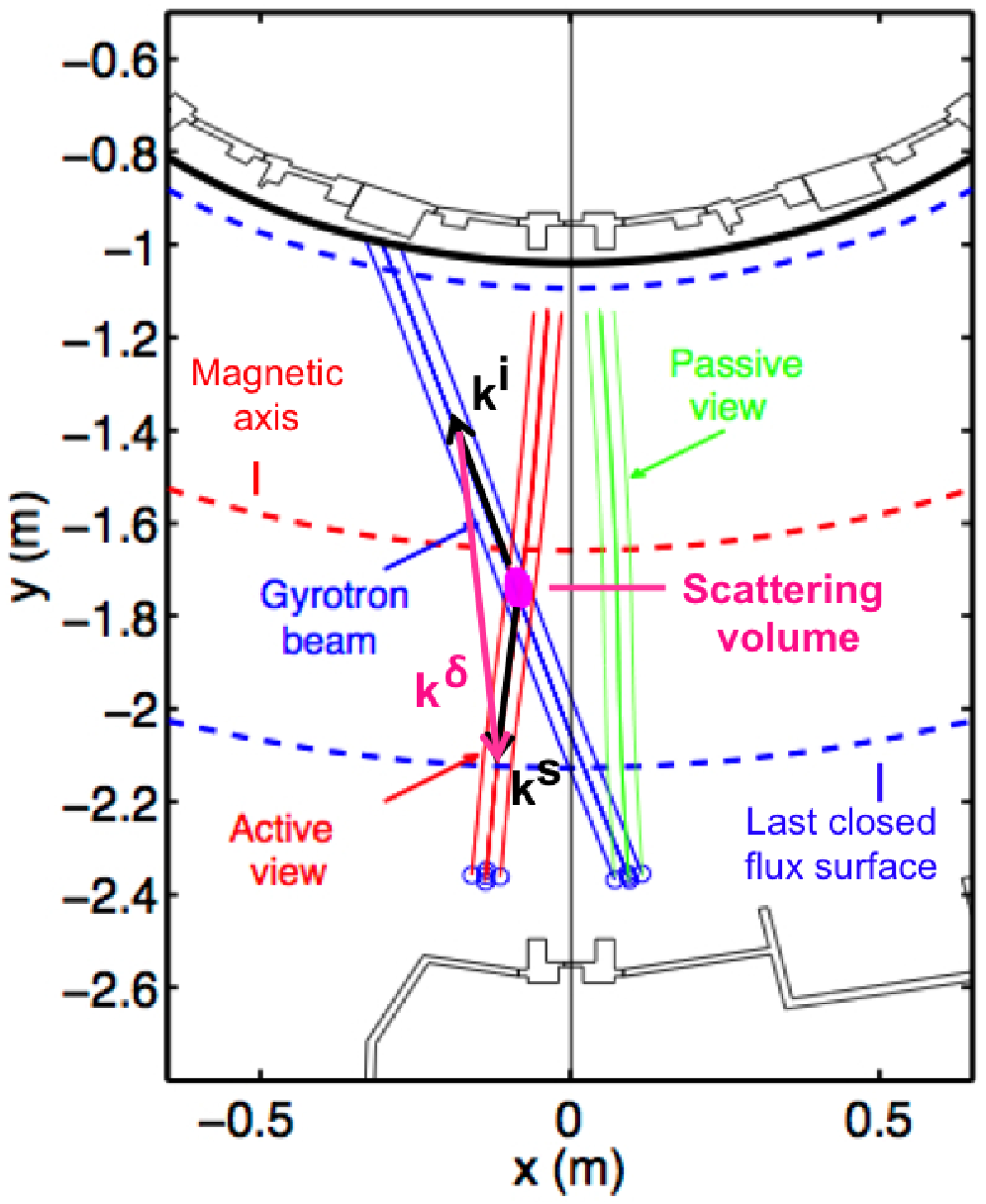}
\caption{CTS geometry in the (left) poloidal and (right) toroidal plane for ASDEX
  Upgrade discharge 29600
  discussed below. The gyrotron probe beam (blue) overlaps with the
  receiver view (red) in the scattering volume (magenta). For
  this viewing geometry, the resolved fluctuation
  vector ${\bf k}^\delta$ makes an angle of $107^\circ$ to the static
  magnetic field.} 
\label{fig,geom}
\end{figure}

We have previously operated a CTS system at the
medium-sized TEXTOR tokamak, providing the first routine measurements of
fast-ion populations in magnetically confined plasmas \cite{bind06}.
Here, time-resolved CTS data showed good agreement 
with expectations for classical slowing down 
of the ion velocity
distribution after turnoff of auxiliary (neutral beam injection; NBI) heating 
\cite{skni08}. However, for some discharges comparison of CTS data with predictions from transport codes 
showed discrepancies that could potentially be explained by anomalous transport \cite{dimo11a}. 

In this paper we describe the first
results obtained with a new dual-receiver CTS setup at ASDEX Upgrade (AUG), including 
preliminary results on fast-ion dynamics.
The CTS system at AUG employs a gyrotron which generates
mm-wave probing radiation at $\nu^i=105$~GHz with a typical probing power of $\sim 600$~kW. 
Scattered radiation is directed by steerable mirrors into antennas which feed a heterodyne receiver system \cite{furt12}.
In the previous setup, a single receiver was used for detecting the scattered
radiation \cite{meo08,meo10,sale10}. By using on/off modulation of the gyrotron probe beam with
2-ms on-times, the background signal, dominated by electron cyclotron
emission, was extracted during gyrotron-off periods where there is no CTS
radiation. However, background-subtracted spectra frequently indicated
the presence of a spurious, residual background signal not expected by CTS theory.
Although the spectral power
density (SPD) of this signal remains at the few eV level, it can still have an
impact on the interpretation of CTS fast-ion signals which have typical SPD
$\approx 1$--20~eV. Moreover, the first analyses
of 1D fast-ion velocity distributions inferred with CTS at AUG showed 
 some quantitative discrepancies between data and predictions by numerical simulations, despite
reasonable qualititative
agreement regarding salient features in the spectra
\cite{meo10,sale10}. These discrepancies are plausibly related to the presence of this
spurious background signal. 

While experimental characterization of the spurious signal is underway,
a model for its dependence on viewing geometry and plasma parameters is not yet in place. Hence,
to  help account for this signal, and more generally to provide
enhanced flexibility in the CTS setup, the CTS receiver at TEXTOR was recently 
transferred to AUG as a "passive-view" receiver system. The viewing direction of this receiver does not
intersect the probe beam (see Fig.~\ref{fig,geom}) and hence does not measure CTS. However, the receiver
has the same poloidal viewing geometry and sees the
same spurious signal as the "active-view" receiver. Data from the new passive view can therefore be used to characterize the residual non-CTS signal seen by the active-view receiver.
As explained, the resulting improvement in the background estimation is of particular importance
for the interpretation of fast-ion CTS data.

\section{First CTS results on fast-ion dynamics at ASDEX Upgrade}

As a first application of the new dual-receiver background approach, we
here present results obtained for the fast-ion populations in AUG discharge 29600. 
This was a standard H-mode discharge with $B_t =
2.55$~T, $I_p=0.8$~MA, and central electron density $n_e=6.6\times
10^{19}$~m$^{-3}$. The scattering volume was placed
near the centre of the plasma, at $(R,z ) = (1.67\mbox{m},
-0.10\mbox{m})$ as estimated from ray-tracing. The CTS temporal and radial spatial resolution for this discharge were 5~ms and 6~cm, respectively.
The angle between the resolved plasma fluctuations and the toroidal
B-field was $\angle({\bf k}^\delta,B)= 107^\circ$ (see Fig.~\ref{fig,geom}).
Co-current neutral beam heating from two sources with similar
injection geometry, Q3 (60~keV injection energy) and Q8
(93~keV), was employed both independently and together during the
discharge to study the slowing down of fast ions in the
plasma. 

\begin{figure}
\includegraphics[width=10.9cm,height=4.75cm]{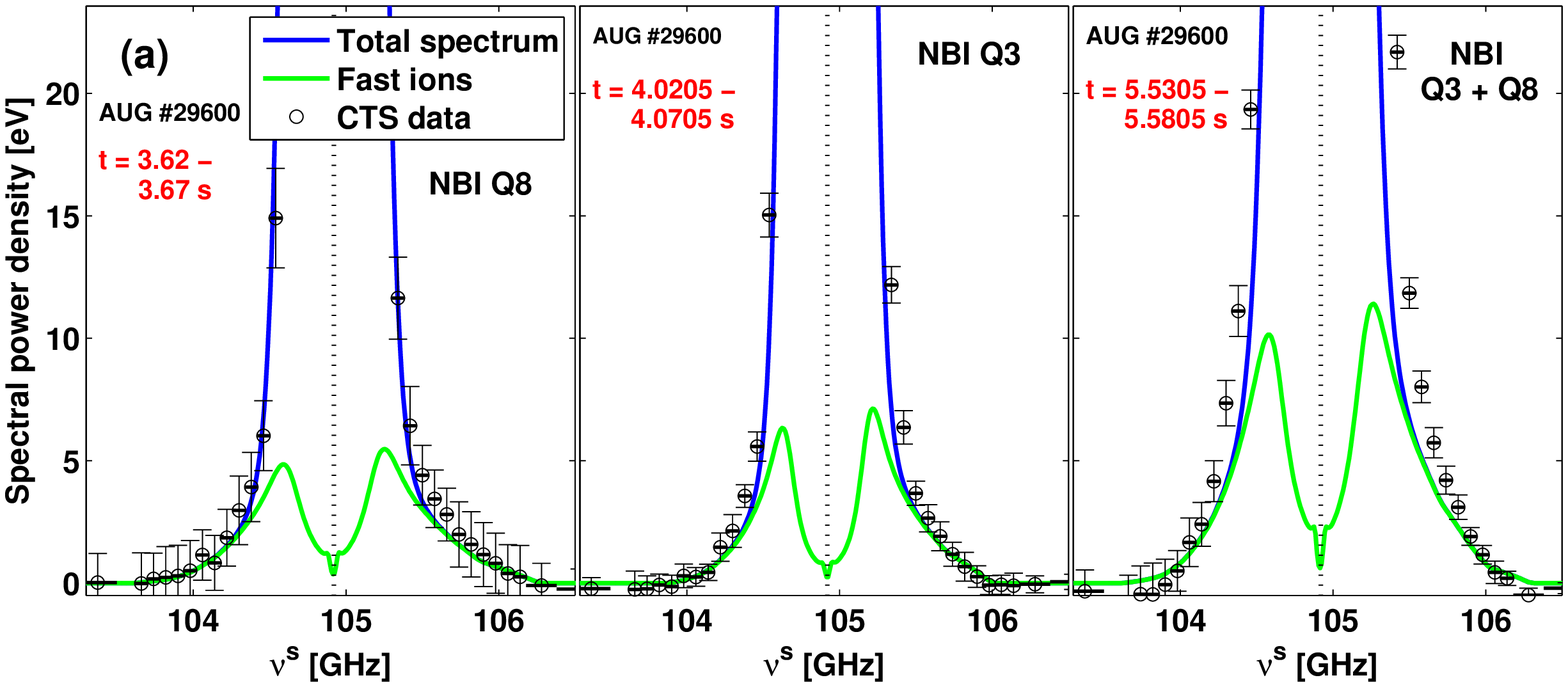}
\includegraphics[width=5.6cm,height=4.7cm]{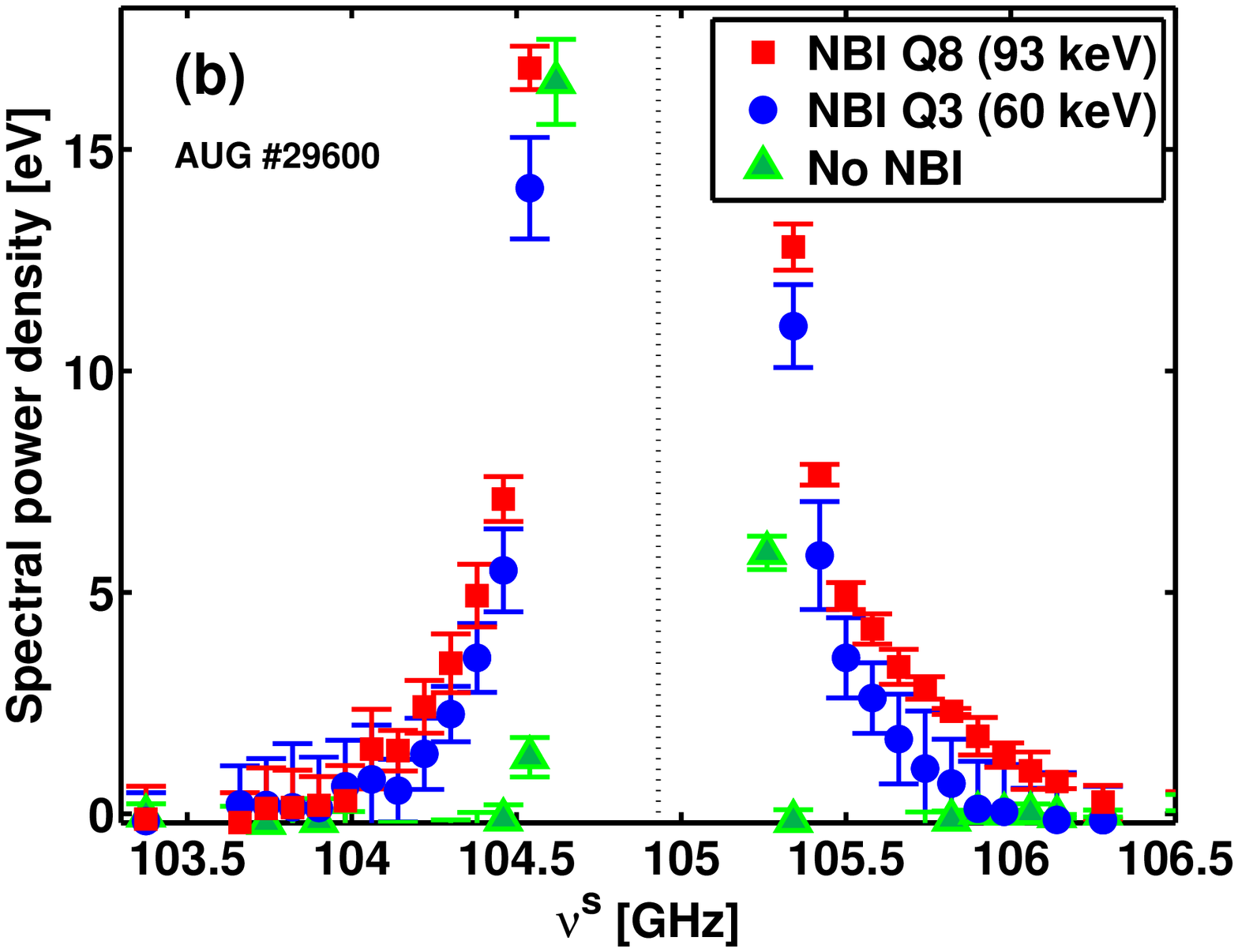}
\caption{({\em a}) CTS spectra  obtained at the end of three separate NBI heating
  phases, compared to synthetic spectra of the
  total (blue) and fast-ion (green) components. ({\em b}) Comparison of CTS
  spectra during phases with and without one-beam NBI heating.
 The CTS data have been averaged over 10 gyrotron pulses. 
 Dotted line marks $\nu^i$ of the gyrotron beam.}
\label{SPD}
\end{figure}

By combining a forward model of the scattering \cite{bind93,bind96} with the fast-ion velocity
distribution in the scattering volume predicted by the transport
code TRANSP \cite{pank04}, one can generate synthetic CTS spectra for comparison to
the real measurements. Previous such comparisons
showed that salient features of simulation and experiment agree, but inexplicable quantitative 
discrepancies remained, with CTS spectra unexpectedly suggesting the presence of a significant fast-ion population with velocities above the NBI injection velocity \cite{sale10}. For discharge 29600, Fig.~\ref{SPD}{\em a} shows a comparison between synthetic
and measured spectra for three different NBI heating phases. With the improved background
subtraction, predicted and observed spectra agree quantitatively during all three phases. 
As shown in Fig.~\ref{SPD}{\em b}, the improved background subtraction also
enables us, for the first time, to clearly observe the expected broadening of the CTS spectrum with increasing
NBI heating energy.

\begin{figure}
\includegraphics[width=5.5cm,height=4.75cm]{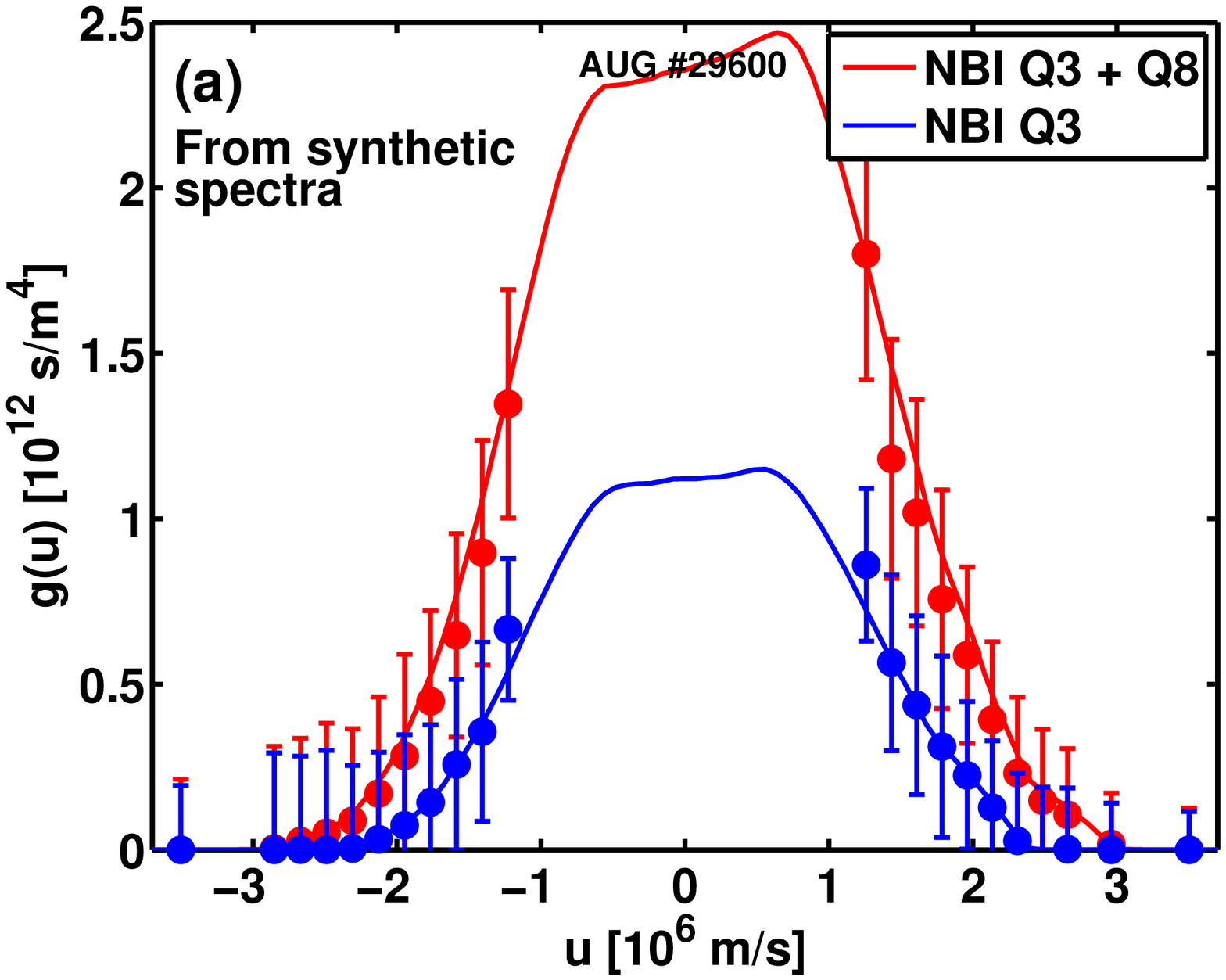}
\includegraphics[width=5.5cm,height=4.75cm]{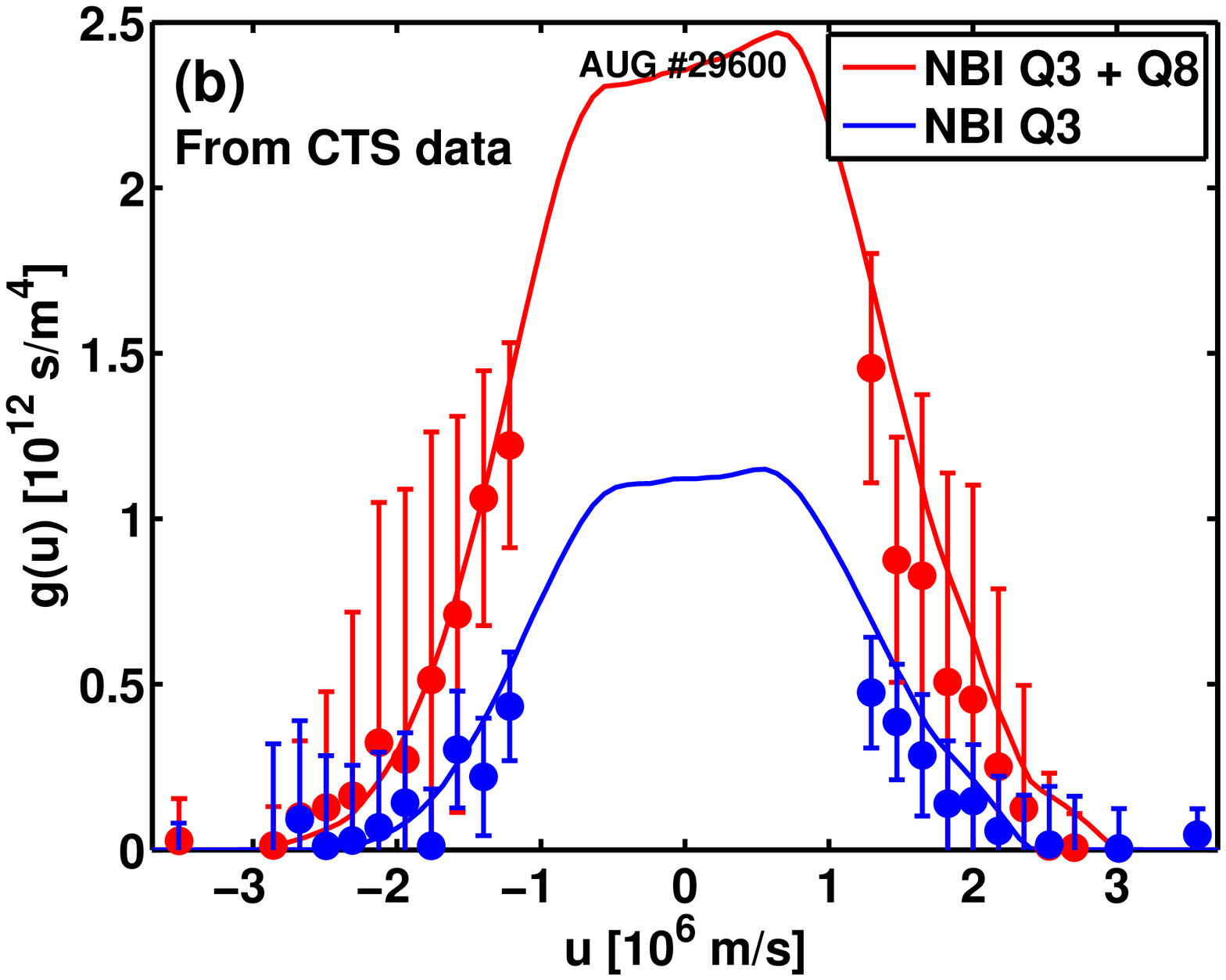}
\includegraphics[width=5.5cm,height=4.70cm]{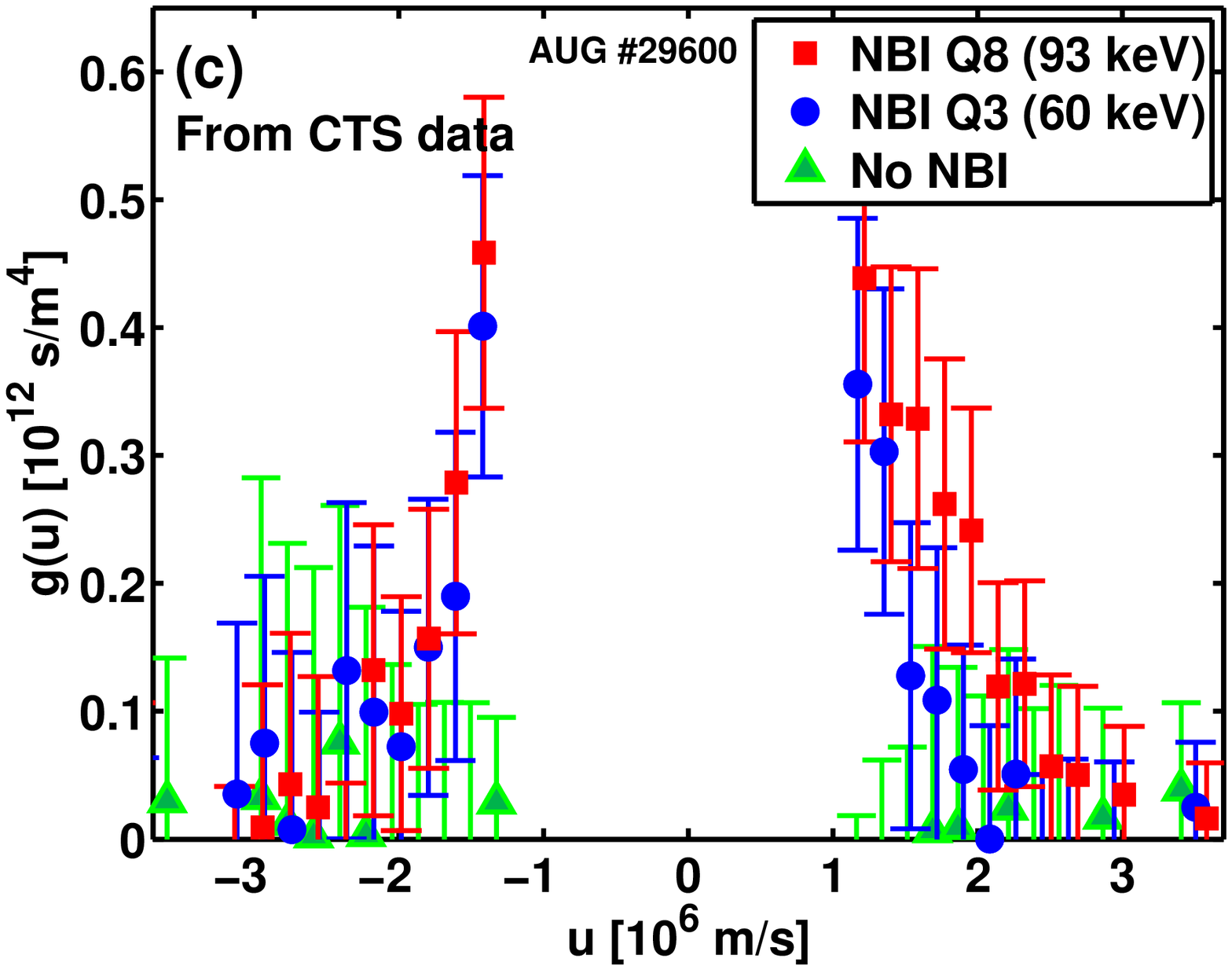}
\caption{({\em a}) Comparison of simulated $g(u)$ from TRANSP (lines) with those
  inferred from the corresponding synthetic CTS spectra (data points) during one- and two-beam
   heating phases. ({\em b}) Simulated $g(u)$ compared to the
  values inferred from real CTS data during the same two NBI phases. ({\em c}) $g(u)$ from CTS data with
 and without one-beam NBI heating.}
\label{vel}
\end{figure}

Fitting the observed CTS spectra with our forward model \cite{bind99}, the 1D (fast) ion velocity distribution $g(u)$,
which is the projection of the fast-ion distribution function onto ${\bf k}^\delta$, can be inferred. As this represents a complicated inverse problem, it is useful to first
verify that the $g(u)$ inferred from synthetic spectra -- i.e.\ those
shown in Fig.~\ref{SPD} -- correctly recover the underlying true $g(u)$
predicted with TRANSP for the given scattering geometry \cite{sale10}. In Fig.~\ref{vel}{\em a} we show that this is indeed the case during both one- and two-beam heating phases for the projected ion velocities 
$u$ beyond the Maxwellian bulk. Fig.~\ref{vel}{\em b} plots the corresponding
comparison for the real CTS data shown in  Fig.~\ref{SPD}, demonstrating, for the first time, good agreement between
theory and experiment during both these heating phases. Finally,  
Fig.~\ref{vel}{\em c} shows that the impact of increasing NBI heating energy
can be discerned in 1D velocity space too, as heating with Q8 produces a broader velocity distribution than with Q3, and as no supra-thermal ions are present when NBI is off.

\begin{figure}
\includegraphics[width=7.4cm,height=5.3cm]{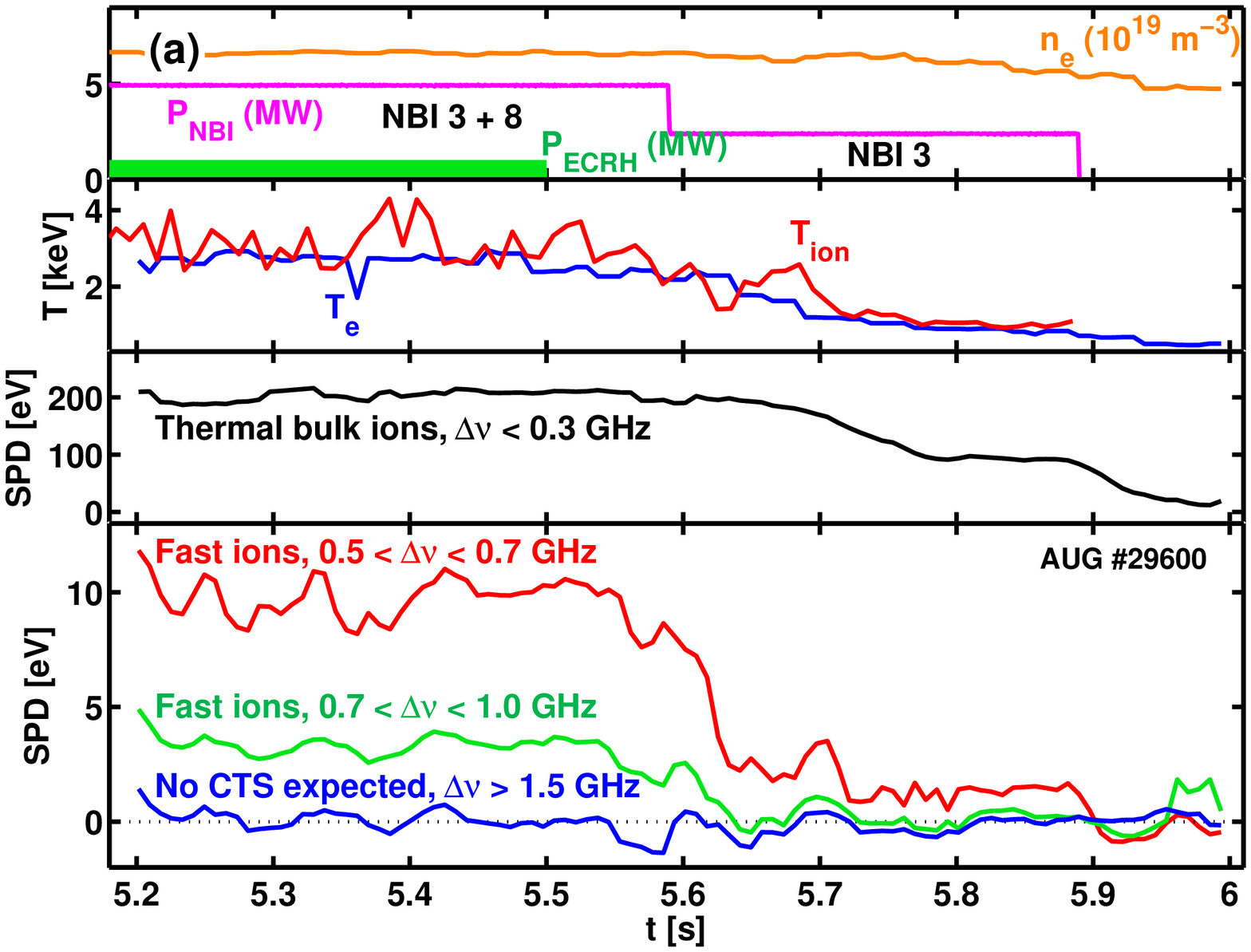}
\includegraphics[width=8.1cm,height=5.4cm]{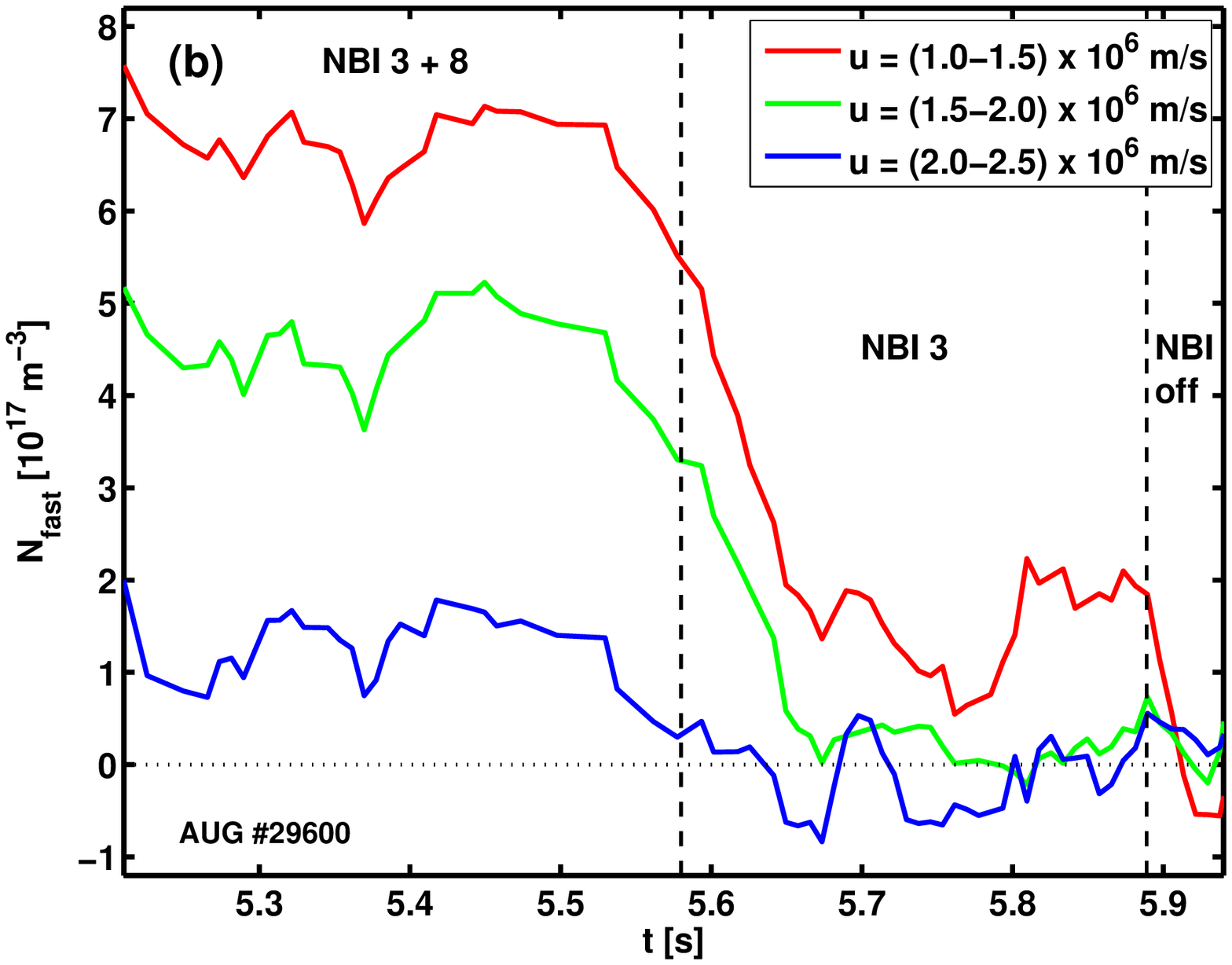}
\caption{({\em a}) Time traces of AUG discharge 29600, showing NBI and ECRH heating
  power, electron density and temperature from Integrated Data Analysis (IDA), ion temperature
  from charge exchange recombination spectroscopy, and
  mean SPD in various frequency ranges $\Delta \nu = |\nu - \nu^i|$ as labelled.
 ({\em b}) Preliminary partial fast-ion densities derived in three 1D velocity intervals.} 
\label{FI}
\end{figure}

These new results represent a major breakthrough for CTS at AUG and open up for studies
of fast-ion dynamics in the present discharge. 
In Fig.~\ref{FI}{\em a} we investigate the slowing down of fast ions after 
auxiliary heating is switched off, by plotting the CTS signal within selected frequency ranges 
$\Delta \nu$ that cover bulk and fast ions. 
Once heating with ECRH and NBI Q8 is switched off at $t\simeq 5.5$ and 5.6~s, respectively, both $T_e$ and $T_{\rm ion}$ drop from 2.5 to 1.0~keV, and the SPD for bulk ions at frequencies outside the rejection frequencies of the notch filters declines. Ions with $0.5<\Delta \nu<0.7$~GHz are predominantly 
supra-thermal and also exhibit a drop but retain an average SPD $>0$ until Q3 is also turned off. Even larger frequency
shifts (0.7--1.0~GHz) are observed only during two-beam heating. No
CTS signal is expected nor observed for $\Delta \nu > 1.5$~GHz. With the low SPDs involved, 
accurate background subtraction is clearly key to identifying 
these trends. 

In Fig.~\ref{FI}{\em b}, we integrate the inferred $g(u)$ over three 1D fast-ion velocity intervals, 
to derive associated partial fast ion densities $N_{\rm fast}$ \cite{skni11}.
All three velocity ranges are represented during the two-beam heating phase, but 
once Q8 is switched off, the faster ions are slowed down,
and $N_{\rm fast}$ relaxes to a new quasi-steady state in $\Delta t\approx 100$~ms. 
Turning off Q3 as well leads to complete absence of a fast-ion CTS signal.
These results agree with expectations from Figs.~\ref{SPD} and 
\ref{vel}, and demonstrate
that CTS studies of fast-ion dynamics at ASDEX Upgrade are now feasible.

\section{Summary and outlook}\label{sec,outlook}

An operating scenario for CTS has now
been established at ASDEX Upgrade, relying on a new  
dual-receiver background subtraction technique.
Application of this technique is especially important for deducing properties of fast
ions, for which the CTS signal is generally relatively low. With this
technique, the inferred properties of fast-ion populations are, for the first time, 
in good quantitative agreement with theoretical expectations
during both one- and two-beam neutral beam injection and when no beams are injected. This represents a breakthrough for CTS at AUG and enables the  first CTS studies of fast-ion dynamics at this machine.
In the short term, our next step will be to provide measurements in multiple scattering volumes using the
new background subtraction technique. Furthermore, we intend to replace the digital
2-ms on/off modulation of the probing gyrotron by 3~kHz analog
modulation, in order to provide better time resolution than the current
$\geq 4$~ms. In the longer term, we anticipate that the
results obtained with the new dual-receiver 
setup could be important for the design of CTS systems in other
fusion experiments, where determination of 
fast-ion distribution functions throughout the plasma is of central importance.

\begin{theacknowledgments}
We thank Henrik Bindslev for fruitful discussions. This work, 
supported by the European Communities under the contract of Association
between Euratom and DTU, was partly carried out within the framework of the
European Fusion Development Agreement. The views and opinions expressed herein
do not necessarily reflect those of the European Commission.
\end{theacknowledgments}

\bibliographystyle{aipproc}   

\end{document}